# GNN-Based Deep Surrogate Modeling of Knee Contact Mechanics: Generalizing Neuromuscular Control Patterns Across Subjects

Zhengye Pan[a,b], Jianwei Zuo[a,b], Jiajia Luo[* a,b]

[a] Biomedical Engineering Department, Institute of Advanced Clinical Medicine, Peking University, Beijing, China

[b] Institute of Medical Technology, Peking University Health Science Center, Peking University, Beijing, China

[*] Corresponding author

**Abstract**

**Background:**

Accumulation of abnormal contact stress is a primary biomechanical driver of acute meniscal tears and chronic osteoarthritis. While Finite Element Analysis (FEA) provides the necessary fidelity to quantify these injury-inducing loads, its high computational cost precludes clinical utility. Emerging deep surrogate models promise real-time assessment but suffer a critical blind spot: they predominantly focus on learning anatomical variations, largely overlooking the neuromuscular control patterns. These dynamic, subject-specific motor strategies fundamentally dictate potentially injurious stress distributions inside the knee.

**Methods:** This study investigates the generalization capability of the topology-aware MeshGraphNet regarding cross-subject neuromuscular control patterns under fixed anatomical conditions. We constructed a dataset using gait data from nine subjects via an OpenSim-FEBio co-simulation platform. The MGN was compared against a structure-agnostic Node-wise MLP using a rigorous grouped 3-fold cross-validation on unseen subjects.

**Results:**

The MGN demonstrated superior fidelity, achieving a correlation of 0.94 with ground truth (vs. 0.88 for MLP). In contrast to the MLP, which exhibited the "peak shaving" defect common in deep learning, MGN significantly reduced peak-stress prediction errors and achieved higher spatial overlap in high-risk regions. This indicates that MGN effectively captured the non-local force-transmission pathways unique to each subject's movement strategy.

**Conclusion:** By mimicking the propagation of physical stress through message passing, MGN successfully decodes the heterogeneity of human neuromuscular control, even under fixed anatomy. This establishes GNNs as robust clinical tools capable of identifying functional injury risks that are invisible to purely geometry-based surrogate models.

# 1 Introduction

As an important internal structure of the knee joint, the meniscus performs core functions, including load transmission, shock absorption, and maintenance of joint stability during human movement. However, these roles also subject the meniscus to abnormal mechanical environments. Epidemiological and biomechanical evidence indicates that the accumulation of abnormal contact stress distributions during movement is a primary factor in the development of acute meniscal tears and in the progression of chronic osteoarthritis (OA) [1,2]. Therefore, precise quantification of the meniscus's internal stress field during dynamic motion is clinically valuable for assessing injury risk, planning personalized surgery, and monitoring rehabilitation.

Currently, the mainstream approach to obtaining the mechanical state of the knee joint internal tissues during movement is to couple multibody inverse dynamics with FEA. In this approach, inverse dynamics uses motion capture data to derive joint boundary conditions, and FEA solves for the internal stress-strain states under constraints of complex geometry and nonlinear material properties [3]. However, although this workflow can provide high-precision physical field data, the iterative solution of high-dimensional equation systems and the construction of high-fidelity meshes require substantial computational resources and time. This currently limits the method to small-sample, retrospective basic research and makes it challenging to meet clinical needs for rapid mechanical evaluation across subjects and across postures [4,5].

To address the efficiency issue in traditional numerical simulations, many studies have introduced deep surrogate models to learn the nonlinear mapping between inputs (boundary conditions, geometric parameters) and outputs (stress and strain fields), thereby replacing FEA solutions at near-real-time speeds [6,7]. In particular, MeshGraphNets (MGN), based on graph neural networks (GNNs), can simulate the propagation of physical quantities between nodes of a discretized mesh via a message-passing mechanism, accurately approximating FEA-derived stress fields across different mesh topologies [8]. This strong cross-geometry generalization capability and advantage in handling complex mesh topologies have made it a cutting-edge surrogate model in computational biomechanics, especially in fluid domains such as hemodynamics [9,10].

However, for musculoskeletal solid mechanics, especially knee mechanics, mesh topology only defines the static geometric boundaries of the problem, whereas neuromuscular control governs the dynamic physical evolution of the system [11]. During movement, the mechanical loading within the knee joint depends on highly individualized, time-varying joint postures, joint contact forces, and muscle contraction forces, which arise from dynamic interactions among the central nervous system, musculoskeletal structures, and the external environment, reflecting each individual's neuromuscular control pattern [12,13]. This implies that knee joint FE surrogate models need not only the ability to handle topology, but also the capability to learn this high-dimensional "control–mechanics" manifold. Current studies on joint surrogate models often focus on generalization over geometric variability while simplifying complex dynamic boundary conditions to static or idealized conditions, and few have investigated the potential of MGN to capture heterogeneous neuromuscular control patterns across subjects [14,15].

This study aims to investigate the generalization capability of MGN, which has topology-aware advantages, under a fixed anatomical structure in the context of cross-subject neuromuscular control patterns. We constructed a knee joint finite element dataset driven by heterogeneous gait data from nine subjects. We used a generic geometric model as a baseline to eliminate anatomical differences, allowing the surrogate model to focus on learning the control–mechanics mapping. For evaluation, we employed a grouped 3-fold cross-validation strategy instead of the traditional leave-one-out approach, forcing the model to be tested on completely unseen subject data to examine its generalization under strict conditions. On this basis, we systematically compared MGN with a node-wise multilayer perceptron (Node-wise MLP) that lacks topology awareness to verify the core hypothesis. Even with a fixed-mesh topology, MGN's explicit message-passing mechanism is significantly superior to the non-structured baseline model in capturing non-local force transmission induced by different neuromuscular strategies, thereby providing a theoretical basis for developing clinical surrogate models with dual robustness in physics and control.

## 2 Methods

### 2.1 Dataset Construction

2.1.1 Subject Recruitment and Data Collection

Nine healthy adult male subjects (height $176.41 \pm 4.6$ cm, weight $73.83 \pm 6.9$ kg, age $23.51 \pm 4.73$ years) were recruited for this study. Inclusion criteria were the absence of knee pain, no history of knee surgery, and no lower limb movement dysfunction. All subjects were informed of the study protocol and purpose and provided written informed consent.

For data acquisition, kinematic and ground reaction force (GRF) data were collected using a Vicon 3D motion capture system (200 Hz, V5, Oxford, UK) and a Kistler 3D force platform (1000 Hz, model 9286AA, Winterthur, Switzerland), respectively. Marker placement followed the protocol of Pan et al. [16]. A single-beam infrared timing system was placed 1 m from the center of the force platform to monitor each subject's running speed. After a warm-up, each subject performed five running trials at 2.5 m/s, and the three trials with speed closest to the target and with complete foot contact on the force platform were selected for further analysis. We defined initial contact as the moment when the vertical ground reaction force (vGRF) exceeded 20 N, and toe-off as when the vGRF fell below 20 N. The time between initial contact and toe-off was defined as the stance phase [17].

2.1.2 Data Processing

Marker trajectories and GRF data were filtered using a fourth-order zero-lag low-pass Butterworth filter with a cutoff of 10 Hz [18]. A full-body musculoskeletal model (with the knee joint having three degrees of freedom: flexion-extension, internal-external rotation, and abduction-adduction) based on Rajagopal et al. [19] was used in OpenSim (v4.5, Stanford University, USA) to perform inverse kinematics and joint reaction force analysis. The 3D knee joint posture was represented by XYZ Euler angles in the tibial coordinate system, and joint reaction forces were decomposed in the same joint coordinate system. The knee joint reaction force includes the combined contributions of muscle tensions crossing the joint, ligament constraint forces, and the external ground reaction force at the joint center, and it is a key physical quantity characterizing cross-subject neuromuscular control patterns [20,21].

2.1.3 Finite Element Simulation and Dataset Generation

To eliminate anatomical variability as a confounding factor, we used the OpenKnee(s) generic finite element mesh as a fixed anatomical baseline. The knee joint angles computed by OpenSim were applied as rigid body kinematic constraints on the femur and tibia, and the joint reaction forces were applied as equivalent loads at the distal femoral reference node. A kinematics-driven quasi-static analysis was performed using FEBio (2.10, University of Utah, USA) with a time step of 0.01 s. To ensure the biofidelity of the mechanical responses of the knee joint structures, the menisci and ligaments were modeled as transversely isotropic Mooney–Rivlin hyperelastic materials to capture the anisotropic behavior of collagen fiber reinforcement (ligaments were loaded with pre-strain), and the articular cartilage was modeled as an isotropic Mooney–Rivlin hyperelastic material.

Each quasi-static timestep was treated as an independent physical state and encapsulated as a graph sample $G=(V, E)$. Here, $V$ and $E$ denote the sets of mesh nodes and edges, respectively. The meniscal elements serve as graph nodes (their centroid coordinates form each node's geometric features). Each model input $X \in R^{10}$ consists of (1) the nodal centroid coordinates in the reference configuration (fixed geometry) and (2) a global driving vector (sine/cosine–encoded joint angles and loads) representing the time-varying control pattern, broadcast to all nodes. Element-wise von Mises stresses were first transformed using $\log(1 + x)$, and then z-score normalized using the mean and standard deviation computed over the entire dataset.

## 2.2 Deep Surrogate Model

### 2.2.1 Network Architecture

Based on the classic MGN framework, this study adaptively modified the network structure to suit the characteristics of musculoskeletal solid mechanics [8]. The overall network still employs an "encode-processor-decode" graph neural architecture. Still, it explicitly incorporates gating mechanisms and global modulation during message passing to accommodate the complex nonlinear mechanical responses of biological tissues, as illustrated in Figure 1.

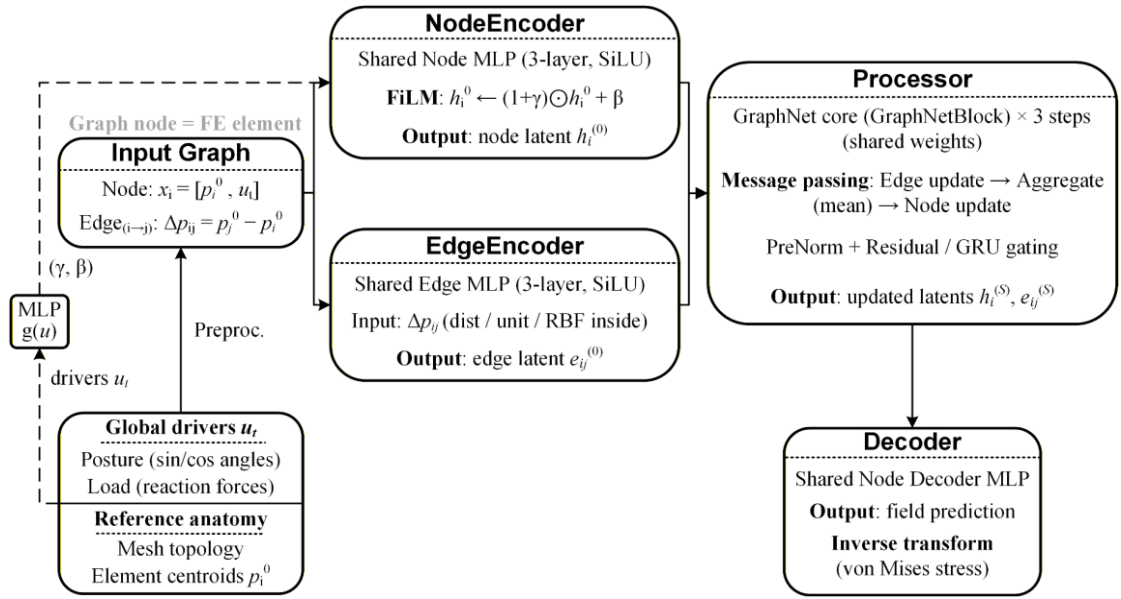

Figure 1. Schematic of the MGN surrogate model architecture

The encoder maps the input node features $X$ and edge features $E$ to high-dimensional latent vectors $h_i^{(0)}$ and $e_{ij}^{(0)}$. To enhance the model's sensitivity to small geometric deformations, we additionally incorporate radial basis function (RBF) distance expansions into the edge features. Meanwhile, the encoder adopts the FiLM (Feature-wise Linear Modulation) mechanism, using the global driving variables (joint angles and loads) to dynamically modulate the node feature distribution [22].

The core processor module consists of three stacked graph network blocks, each performing a message-passing operation at step *k*. In each step, a multilayer perceptron aggregates messages from a node's neighborhood and updates the node state.

$$e_{ij}^{(k+1)} = \phi^e \left( h_i^{(k)}, h_j^{(k)}, e_{ij}^{(k)} \right) \tag{1}$$

$$h_i^{(k+1)} = \phi^v \left( h_i^{(k)}, \sum_{j \in \mathcal{N}(i)} e_{ij}^{(k+1)} \right) \tag{2}$$

Here, $\phi^e$, $\phi^v$ denote multilayer perceptrons with residual connections, and $\mathcal{N}(i)$ represents the neighborhood of node *i*. This iterative process is mathematically analogous to the local transmission and equilibrium of stress waves and energy in a continuous medium mesh. Unlike the original MGN, which uses only residual connections, we incorporate a gated recurrent unit (GRU) gating mechanism in the message update step [23]. Mathematically, this means that the node state update depends not only on the aggregation of messages from neighbors but is also modulated by the current hidden state, enabling more effective retention of long-range dependencies and alleviating the oversmoothing problem during iterations. Finally, the decoder maps the evolved node states back to the physical scalar space and outputs the predicted element-wise von Mises stress.

In addition, we implemented a node-wise multilayer perceptron (Node-wise MLP) as a baseline model. The architecture comprises four fully connected layers; each hidden layer contains 256 neurons and uses the SiLU nonlinearity. This Node-wise MLP completely discards the mesh topology (i.e., edge information) and treats each node in the finite-element mesh as an independent sample in space. During both training and inference, the model takes only each node's local geometric features and global driving variables as inputs and performs pointwise stress regression. By intentionally eliminating inter-nodal physical information exchange, this baseline serves as an ideal control for evaluating the benefits of the MGN.

2.2.2 Training and Inference

The model training aimed to minimize the difference between the predicted stress field and the finite element ground truth. The model performed regression and loss calculation in the log(1+x) transformed and standardized stress space; after training, the predictions were inverse-transformed back to the original stress scale. Since the mesh data may contain filler nodes in non-meniscus regions, we adopted a masked mean squared error (MSE) loss, computing the regression error only for nodes of actual tissue to ensure that gradient updates focused on the region of interest. The AdamW optimizer was used with an initial learning rate of $1\times10^{-3}$ and a ReduceLROnPlateau scheduler: if the validation loss did not decrease for 10 consecutive epochs, the learning rate was automatically reduced by 50% to stabilize convergence. All models were trained on two NVIDIA RTX TITAN GPUs (batch size = 1; epochs = 100).

To configure the model hyperparameters, we first conducted a set of lightweight ablation experiments. The results indicated that, for the present musculoskeletal solid-mechanics mapping task, the prediction accuracy of the MGN saturated when the hidden dimension was set to 32 and the number of message-passing steps to 3. Accordingly, we adopted this lightweight MGN configuration ($D_{hidden}$ = 32, K = 3) to achieve an optimal balance between efficiency and accuracy. In contrast, the baseline MLP was assigned a larger capacity ($D_{hidden}$ = 256) to ensure a fair comparison.

To rigorously assess the model's ability to generalize to unseen neuromuscular control patterns, we adopted a grouped 3-fold cross-validation scheme. The nine subjects were partitioned by ID into three folds: Fold 1 ($P_1$–$P_3$), Fold 2 ($P_4$–$P_6$), and Fold 3 ($P_7$–$P_9$). In each fold, the model was trained on data from two subject groups and tested on the remaining group of entirely unseen subjects, thereby eliminating any possibility of subject-level data leakage. Early stopping was applied in every fold, with training terminated if the validation loss failed to improve for 30 consecutive epochs.

2.2.3 Comparative Metrics

For global engineering accuracy, we used root-mean-square error (RMSE) and mean absolute error (MAE) to quantify overall prediction error, and introduced normalized RMSE (nRMSE) to eliminate dimensional effects. Suppose at a given time step *t*, the total number of nodes is *N*, with the finite element ground-truth stress for node *i* being $y_i$ and the model's predicted stress $\hat{y}_i$.

$$\text{RMSE} = \sqrt{\frac{1}{N}\sum_{i=1}^{N}(\hat{y}_i - y_i)^2} \tag{3}$$

$$\text{MAE} = \frac{1}{N}\sum_{i=1}^{N}|\hat{y}_i - y_i| \tag{4}$$

$$\text{nRMSE} = \frac{\text{RMSE}}{\max_j(y_j)} \tag{5}$$

where max(*y*) is the peak ground-truth stress at that time step. The nRMSE directly reflects the predicted error relative to the instantaneous maximum load.

Considering that meniscal injuries often occur in regions of high stress concentration, fidelity in key areas is critical. We define the peak relative error ($RE_{max}$) and the 95th-percentile relative error ($RE_{P95}$) to specifically test whether the model exhibits the "peak shaving" phenomenon common in deep surrogate models:

$$RE_{max} = \frac{|max_i(\hat{y}_i) - max_i(y_i)|}{max_i(y_i)} \tag{6}$$

$$RE_{P95} = \frac{|P_{95}(\hat{y}) - P_{95}(y)|}{P_{95}(y)} \tag{7}$$

where $P_{95}(\cdot)$ denotes the 95th percentile. These metrics capture the model's prediction bias in the most dangerous regions of the meniscus.

To verify spatial consistency between the predicted and ground-truth stress fields, we computed the Pearson correlation coefficient (*r*) at the node level. We introduced the Dice coefficient and intersection-over-union (IoU) to evaluate the overlap of "high-risk regions." These high-risk regions are defined as the sets of nodes with stress values above the 90th percentile ($S = \{i \mid y_i > P_{90}(y)\}$). The Dice coefficient and IoU are given by:

$$r = \frac{\sum(\hat{y}_i - \bar{\hat{y}})(y_i - \bar{y})}{\sqrt{\sum(\hat{y}_i - \bar{\hat{y}})^2}\sqrt{\sum(y_i - \bar{y})^2}} \tag{8}$$

$$\text{Dice} = \frac{2|S_{true} \cap S_{pred}|}{|S_{true}| + |S_{pred}|} \tag{9}$$

$$\text{IoU} = \frac{|S_{pred} \cap S_{true}|}{|S_{pred} \cup S_{true}|} \tag{10}$$

where *A* and *B* are the sets of nodes identified as high-risk in the predicted and ground-truth stress fields, respectively. These metrics have clear clinical significance for validating the model's ability to localize potential injury regions via stress localization accurately.

**2.3 Statistical Analysis**

All comparisons between MGN and the Node-wise MLP for the above metrics were made using paired Wilcoxon signed-rank tests. All tests were two-sided with a significance level of α = 0.05. Each metric is reported as mean ± standard deviation. All statistical analyses were performed in Python (v3.9.13, Delaware, US).

# 3 Results

## 3.1 Overall Node Stress Prediction Accuracy Comparison

Table 1 and Figure 2 show that even with fixed anatomy, the MGN model's overall stress prediction performance is significantly superior to that of the node-wise MLP of equal capacity. As shown in Table 1, MGN's mean RMSE and MAE are both significantly lower than those of MLP ($p < 0.01$). In Figure 2, the correlation coefficient $r$ between MGN's predictions and the FE ground truth is significantly higher (0.94 vs. 0.88), and the distribution of nRMSE is significantly lower. This implies that MGN fits the overall nodal stress more accurately and with less variation than MLP.

Table 1. Comparison of overall nodal stress prediction errors between the MGN and MLP models

|      | MGN           | MLP           | median_diff | $p$    |
|------|---------------|---------------|-------------|--------|
| RMSE | 0.307 ± 0.074 | 0.431 ± 0.099 | 0.12        | < 0.01 |
| MAE  | 0.101 ± 0.02  | 0.120 ± 0.03  | 0.02        | < 0.01 |

Note: *median_diff* denotes the median, across the nine subjects, of the difference between MGN and MLP.

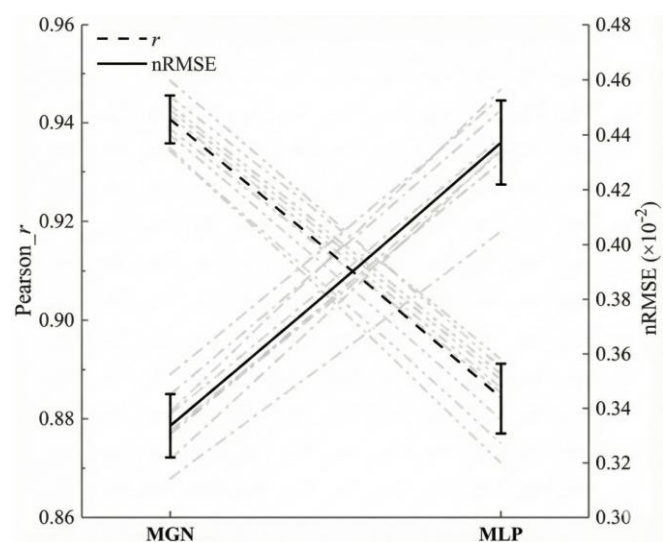

Figure 2. Comparison of the correlation and nRMSE of node stress predictions between MGN and MLP

## 3.2 High Stress Regions and Clinically Relevant Metrics Comparison

Figure 3 indicates that MGN is more reliable at predicting extreme stress values. MGN's relative errors for the stress peak ($RE_{max}$) and the 95th percentile ($RE_{P95}$) are lower than those of MLP, indicating that MGN better captures the highest stress levels. At the same time, MLP exhibits an evident "peak shaving" phenomenon.

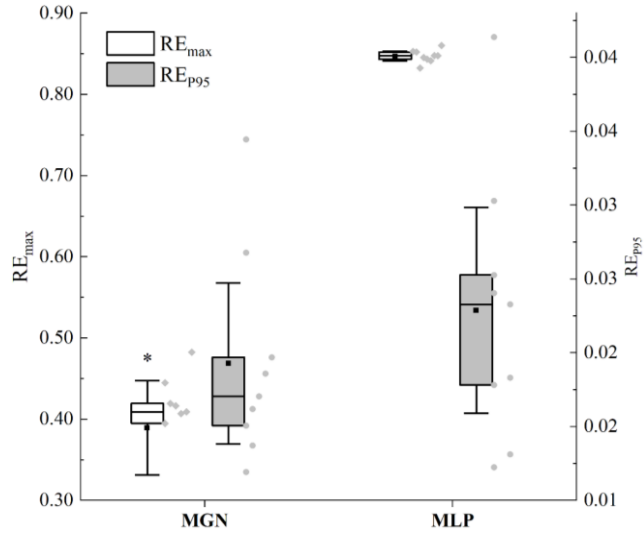

Figure 3. Comparison of the relative errors for stress peak values ($RE_{max}$) and 95[th] percentile ($RE_{P95}$) between MGN and MLP

The spatial consistency of the high-risk regions (stress above the 90th percentile) for the two models is shown in Figure 4. Compared to MLP, MGN achieves significantly higher Dice and IoU values ($p < 0.01$), indicating that its predicted high-stress regions overlap better with the finite element results. In other words, MGN more accurately identifies high-risk regions of the meniscus, whereas MLP may miss or misclassify them.

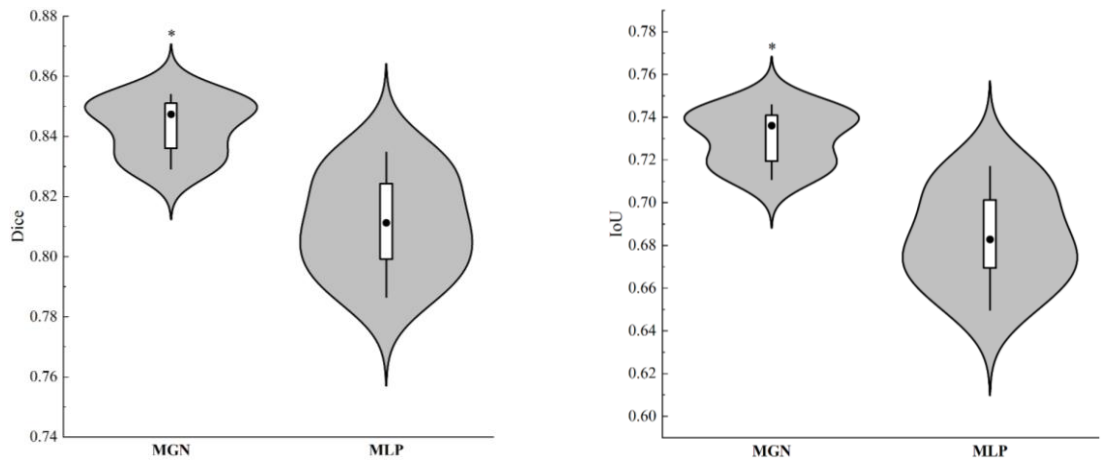

Figure 4. Comparison of Dice coefficient and IoU for high-stress regions between MGN and MLP

## 3.3 Time-Dependent Nodal Stress Prediction Errors During the Stance Phase

Figure 5 shows the time evolution of the prediction error during the stance phase for both models. Overall, MGN's nodal RMSE and peak relative error ($RE_{max}$) remain lower than MLP's throughout the stance phase, indicating that MGN not only accurately predicts stress at individual time points but also more stably tracks the stress time series under complex loading; MLP tends to exhibit larger deviations at peak moments during movement.

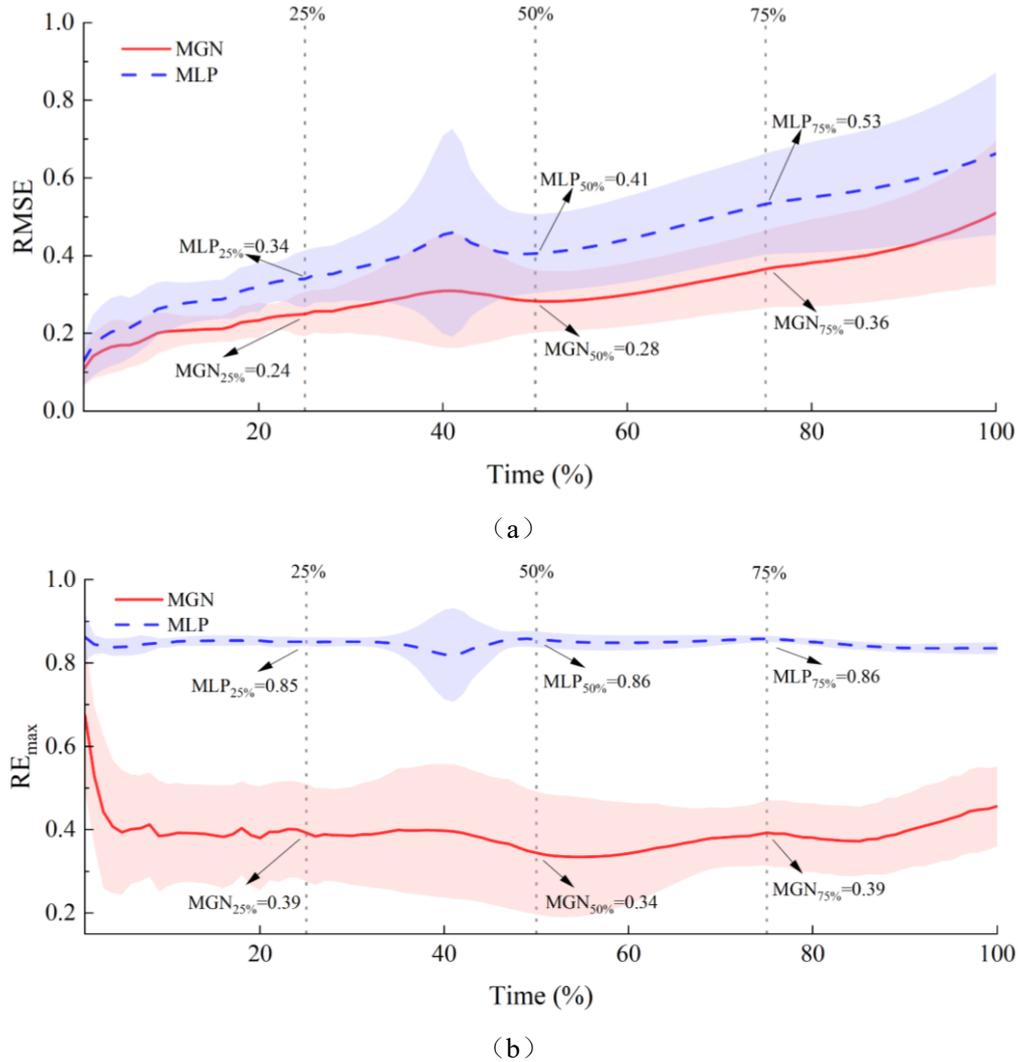

Figure 5. Time-varying characteristics of node stress prediction errors for MGN and MLP during the stance phase: (a) RMSE; (b) $RE_{max}$. Shaded regions denote the standard deviation for MGN (red) and MLP (blue)

Figure 6 presents representative bilateral meniscal stress contour plots for the subject with median RMSE. From the visualization, MGN's predictions closely match the finite element simulation results. At high-stress concentration points such as the posterior horn of the medial meniscus and lateral regions, MGN successfully captures similar high-value (red) areas. In contrast, MLP's predicted map is smoother and cannot accurately reproduce these sharp high-stress patches, often showing shifted locations or underestimated peak intensities. For example, in the high-pressure region at the posterior of the medial meniscus, MGN still shows a clear stress concentration. In contrast, the corresponding area in the MLP map appears lighter in color and reduced in extent. Overall, MGN reconstructs the spatial structure of key stress distributions more accurately and reliably, whereas MLP shows substantial errors in both localization and magnitude.

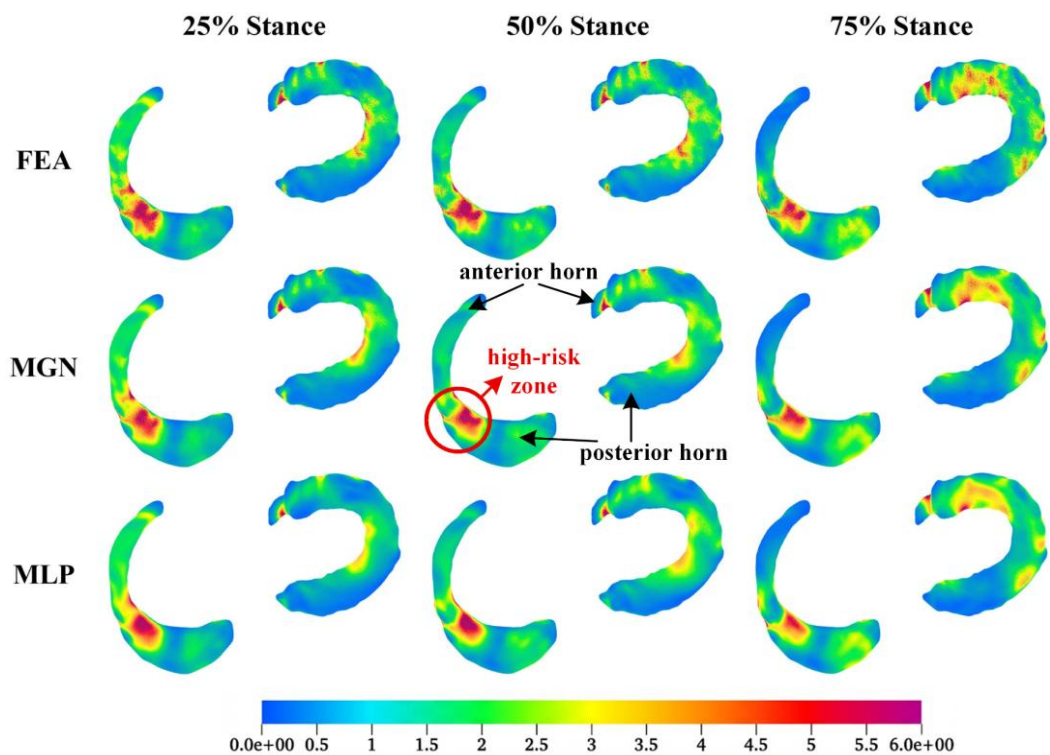

Figure 6. Bilateral meniscal stress (von Mises stress) contour plots for the representative subject with median RMSE

## 4 Discussion

This study compared the performance of MGN and a node-wise MLP on a fixed anatomical mesh, investigating the effect of MGN's structure-aware mechanism on the capture of cross-subject neuromuscular control patterns. The results indicate that even after eliminating differences in mesh topology, MGN significantly outperforms the non-structured MLP baseline in global error control, peak capture, and high-risk region localization. This finding reveals the intrinsic advantage of graph neural networks in modeling complex mechanical responses arising from individualized movement strategies.

Cross-subject heterogeneity manifests not only as differences in the statistical distributions of inputs (e.g., postures and loads), but also as changes in the non-local mechanical propagation characteristics within joint tissues. During movement, each individual's personalized neuromuscular control pattern determines specific knee postures and dynamic contact loads, and these boundary conditions produce completely different stress transmission pathways within the knee joint tissues [24], i.e., complex stress redistribution processes. For example, a specific foot landing strategy may rapidly transmit local impact on the tibial plateau to the opposite side via the meniscus's circumferential fibers [25], forming distal stress concentrations. These long-range dependencies of non-local mechanical propagation constitute core features of the meniscus's dynamic mechanical response [26]. Therefore, cross-subject differences in neuromuscular control strategies essentially manifest as distinct non-local mechanical propagation patterns in knee joint structures, such as the meniscus.

In this context, the limitation of the node-wise MLP is that it reduces the physical field to isolated spatial samples, thereby cutting off topological connections between nodes [27]. Consequently, when faced with differentially propagated stress distributions across subjects, the MLP struggles to reconstruct the correct stress distribution gradients, leading to significant prediction errors in high-stress regions. In contrast, MGN's message-passing mechanism has a natural computational isomorphism with the stress propagation process in continuum mechanics. In each iteration, node features aggregate information from neighbors through edges, which mathematically approximates the diffusion and equilibrium of stress waves or energy in a finite element mesh [8]. When a subject's neuromuscular control pattern changes and alters the mechanical propagation path (for example, shifting from predominantly medial loading to combined medial-lateral loading), MGN can dynamically adjust the message-passing flow via the graph topology, thus accurately capturing these changes in non-local mechanical response. Therefore, even with a fixed anatomical mesh, MGN achieves robust generalization to different neuromuscular control patterns by virtue of its simulation of the physical propagation process.

This cross-subject generalization capability has clear physiological significance. On one hand, neuromuscular control strategies are often important determinants of injury risk and rehabilitation outcomes [28]. For example, under the same external task, different individuals may transfer forces via a "rigid landing" strategy or a "flexible cushioning" strategy, with the former more likely to cause peak stress concentrations in the meniscus and cartilage. MGN can distinguish and reconstruct the non-local mechanical patterns induced by these control strategies even under fixed geometry, suggesting its potential as an evaluation tool for "control-level interventions" to compare how different training or rehabilitation protocols affect joint stress distributions. On the other hand, our findings indicate that, with only a non-structured baseline model, the model tends to capture only the average effect of control inputs on local stress levels and fails to identify dangerous patterns transmitted remotely through complex force chains. This could lead to a systematic underestimation of the effectiveness of high-risk control strategies in practical applications.

It is worth noting that the dataset in this study was constructed under the assumption of quasi-staticity, ignoring the history dependence of neuromuscular control strategies. The current frame-by-frame prediction approach cannot capture dynamic mechanical memory determined by loading history. Future work will introduce a Transformer module into the MGN and use its self-attention mechanism to construct a spatiotemporal graph neural network that aggregates long-term temporal dependencies across gait cycles, thereby testing the model's ability to capture cross-subject non-local mechanical propagation features dynamically.

## 5 Conclusion

This study demonstrates that the generalization ability of deep surrogate models in musculoskeletal solid mechanics should not be limited to adaptability to geometric variation, but should also include internalization of complex mechanical propagation laws. MGN, through its structured message-passing mechanism, successfully captures the cross-subject non-local mechanical propagation features induced by individualized neuromuscular control differences. This finding provides an important theoretical basis for developing clinical biomechanical surrogate models that combine anatomical inclusivity and functional robustness.

## Funding Statement

This study was supported by the National Key R&D Program of China (grant no. 2023YFC2411201); Beijing Natural Science Foundation (grant no. L259081); NSFC General

Program (grant no. 31870942); Peking University Clinical Medicine Plus X – Young Scholars Project (grant nos. PKU2020LCXQ017 and PKU2021LCXQ028); and PKU-Baidu Fund (grant no. 2020BD039).